\documentclass[a4paper,11pt]{article}
\pdfoutput=1 

\usepackage{jheppub} 

\usepackage[T1]{fontenc} 
\usepackage{graphicx}

\title{\boldmath A unified thermodynamic picture of Ho\v{r}ava-Lifshitz black hole in arbitrary space time}

\author[a,1]{Jishnu Suresh,\note{Corresponding author.}}
\author[b]{Tharanath R,}
\author[c]{and V C Kuriakose}

\affiliation[a,b,c]{Department of Physics, Cochin University of Science and Technology,\\Kochi - 682 022, Kerala, India}

\emailAdd{jishnusuresh@cusat.ac.in}
\emailAdd{tharanath.r@gmail.com}
\emailAdd{vck@cusat.ac.in}

\abstract{In this paper, we analyze the complete thermodynamic and phase transition phenomena of a black hole solution in 
Ho\v{r}ava-Lifshitz gravity in arbitrary space time.
Nature of phase transition is studied using geometrothermodynamic
and Ehrenfest's scheme of standard thermodynamics. We analytically check the Ehrenfest's 
equations near the critical point, which is the point of divergence in the 
heat capacity. Our analysis revels that this black hole exhibits a second order
phase transition. }

\begin{document} 
\maketitle
\flushbottom

\section{Introduction}
\label{introduction}

Black hole thermodynamics has been a fascinating topic of study since black holes were identified as thermodynamic objects with both temperature 
and entropy \cite{bekenstein, hawking1}. 
Thermodynamic properties of black holes have been studied during all these years.  During this period, it became well known that the black hole spacetime
can possess
phase structures along with the standard thermodynamic variables like temperature, entropy etc. Hence it causes to believe in the existence of a complete 
analogy between black hole system and non-gravitational thermodynamic systems. Various black hole thermodynamic variables and their properties have been 
extensively studied. In 1983, Hawking and Page \cite{hawking2} discovered the phase transition phenomena 
in the Schwarzschild AdS background. This became a turning point in the study of black hole phase transition. 
After this many studies have been done in this regard \cite{Chamblin1,Chamblin2, Caldarelli, Nojiri, Cai1, Cveti, Carlip, Cai2, Myung1, 
Carter, Cai3, Myung2, Myung3, Koutsoumbas, Cadoni, Liuhaishan, Sahay, Cao, Weishaowen1, Majhi,
Kim1, Tsai, Capela, Kubiznak, Weishaowen2, Eune}. 

Recently geometric method has been identified as a convenient tool to study the thermodynamics and corresponding phase transition
structures of black holes. 
Various investigations are also done by incorporating this idea from information geometry to the study of 
black hole thermodynamics \cite{gibbs,caratheodory,hermann,mrugala1,mrugala2}. 
Riemannian geometry 
to the equilibrium space was first introduced by Weinhold and Ruppeiner. In 1976 Weinhold \cite{weinhold} proposed a metric, as the Hessian of the internal energy, given as $g_{ij}^{W}=\partial_{i} \partial_{j} U \left(S,N^{r} \right)$. Later in 1979,
Ruppeiner\cite{ruppeiner} introduced another metric as the Hessian of entropy, as $g_{ij}^{R}=-\partial_{i} \partial_{j} S \left(M,N^{r} \right)$. 
This Ruppeiner metric 
is conformally equivalent to Weinhold's metric and the geometry that can be obtained from these two methods are related through 
the relation where \cite{mrugala3,salamon1}, 
$ds^{2}_{R}=\frac{1}{T} ds^{2}_{W}$. Since these matrices depend on the choice of thermodynamic potentials, they are not Legendre invariant.
The results obtained with the above two metrics are found to be consistent with the systems like ideal classical gas, multicomponent ideal gas, 
ideal quantum gas, one-dimensional Ising model, van der Waals model etc,  
\cite{jany1,jany2,brody1,brody2,brody3,dolan1,dolan2,janke1,janke2,janke3,johnston}.
But these two metrics fail in explaining the thermodynamic properties and they lead to many puzzling situations. By incorporating the idea of Legendre
invariance, Quevedo et.al \cite{quevedo1,quevedo2,quevedo3} proposed a new geometric formalism, known as the Geometrothermodynamics. The metric structure related with 
geometrothermodynamics can give well explanations for different behaviors of black hole thermodynamic variables. This method seizes the exact phase structure
of black hole systems.

On the other hand, classical thermodynamics can be applied to black hole systems directly to study the phase structure. In classical thermodynamics
first order phase transitions satisfies Clausius-Clapeyron equation, while second order phase transitions satisfies Ehrenfest equations.
Recently Banerjee et al. \cite{banarjee1,banarjee2,banarjee3,banarjee4,banarjee5,banarjee6} developed a new scheme to study the phase 
transition in black holes, based on Ehrenfest equations by considering 
the analogy between thermodynamic variables and black hole parameters. This Ehrenfest scheme provides a unique way to classify the nature of phase 
transitions in black hole systems. If two Ehrenfest relations are satisfied by a black hole system, then the corresponding phase transition can
be identified as second order in nature. Even if it is not second order, we can find the deviation from second order by defining
Ehrenfest's relation as quantified by defining a parameter called  the Prigogine-Defay(PD) ratio \cite{th1,th2,jackle}. Interestingly many 
calculations show that there are 
black hole systems whose phase transitions lie within the bound conforming to a glassy phase transition.

Ho\v{r}ava-Lifshitz theory is an anisotropic and non-relativistic renormalizable theory of gravity at a Lifshitz point, which can be treated as a good 
candidate for the study of quantum field theory of gravity, and it retrieves the Einstein's gravity in the IR limit. Recently its black hole solution and
thermodynamics have been intensively investigated \cite{LMP, Caicaoohta, KS, Nastase, Kofinas, Calcagni, Park1, Wei, Myung_1, Myung_2, Kim, nv1, Park, js1,js2}.
By introducing a dynamical parameter $\Lambda$ in asymptotically $\textmd{AdS}_4$ space-time,
a spherically symmetric black hole solution was first given by Lu et al.\cite{LMP}. In this paper, motivated by all the above mentioned features,
we extract the whole thermodynamic quantities of this  black hole, called Lu Mei Pope (LMP) black hole in arbitrary space time. Also we study the phase structure 
of this solution using geometric methods as well as using the Ehrenfest scheme. 

The paper is organized as follows. In section \ref{thermo}, we discuss the thermodynamics of LMP black hole in HL gravity and we calculate
the equation of state of the black hole system. Using the idea of geometrothermodynamics, the thermodynamics and the peculiar behaviors of thermodynamic 
variables are studied in section \ref{gtd}. In section \ref{ehren}, using the Ehrenfest scheme, the nature of phase transition is discussed. Finally the 
conclusions are quoted in section \ref{Conclusions}.

\section{Review of Thermodynamics of LMP black hole }
\label{thermo}

Ho\v{r}ava used the ADM formalism, where the four-dimensional metric of general relativity 
is parameterised as \cite{adm},
\begin{equation}
 ds_4^2= - N^2  dt^2 + g_{ij} (dx^i - N^i dt) (dx^j - N^j dt) ,
 \label{metric_adm}
\end{equation}
where $N$, $N^i$ and $g_{ij}$ are the lapse, shift and 3-metric respectively.  The ADM decomposition (\ref{metric_adm}) of the
Einstein-Hilbert action is given by,
\begin{equation}
 S_{EH} = \frac{1}{16\pi G} \int d^4x \sqrt{g} N (K_{ij} K^{ij} - K^2 + R -
2\Lambda),
\label{eh_action}
\end{equation}
where $G$ is Newton's constant, $R$ is the curvature scalar and $K_{ij}$ is defined by,
\begin{equation}
 K_{ij} = \frac{1}{2N} (\dot g_{ij} - \nabla_i N_j - \nabla_j N_i).
\end{equation}

The action of the 
theory proposed by Ho\v{r}ava \cite{horava1} can be written as,
\begin{eqnarray}
 S_{HL} &=& \int  dt d^3 x \sqrt{g}N [ \left\{\frac{2}{\kappa^2}(K_{ij}K^{ij} -\lambda  K^2)+\frac{\kappa^2\mu^2(\Lambda_W R -3\Lambda_W^2)}{8(1-3\lambda)}\right\} \nonumber \\
&+&\left\{\frac{\kappa^2\mu^2 (1-4\lambda)}{32(1-3\lambda)}R^2-\frac{\kappa^2}{2w^4} \left(C_{ij}
-\frac{\mu w^2}{2}R_{ij}\right)  \left(C^{ij} -\frac{\mu w^2}{2}R^{ij}\right)\right\} ] ,
\label{shl}
\end{eqnarray}
where $\lambda\,,\kappa\,,\mu\,,w$ and $\Lambda_W$ are constant
parameters, and $C_{ij}$ is the Cotton tensor, defined by,
\begin{equation}
C^{ij}=\epsilon^{ik\ell}\nabla_k\left(R^j{}_\ell
-\frac14R\delta_\ell^j\right).
\end{equation}
Comparing the first term in the (\ref{shl}) with that of general relativity in the ADM
formalism, we can write the speed of light, Newton's constant and the cosmological
constant respectively as,
\begin{eqnarray}
 c=\frac{\kappa^2\mu}{4} \sqrt{\frac{\Lambda_W}{1-3\lambda}}, \qquad
 G=\frac{\kappa^2}{32\pi c} \quad
\textmd{and}  \quad \Lambda=\frac{3}{2} \Lambda_W.
\end{eqnarray}
Now we will look for a static and spherically symmetric solution with the metric,
\begin{equation}
 ds^2= f(r) dt^2-\frac{dr^2}{f(r)}+r^2 d\Omega ^2 .
 \label{line element}
\end{equation}
In the present study we are interested in the solution with the  choice  $\lambda =1$. This will lead to the LMP black hole solution \cite{LMP}, given by,
 \begin{equation}
 f(r)= k- \Lambda_W r^2 - \mathcal{A} \sqrt{\frac{r}{-\Lambda_W}}.
 \label{fofr}
\end{equation}
where $\mathcal{A}$ is an integration constant and is related to the black hole mass as $\mathcal{A}=a M.$ It is interesting to note that
this solution (\ref{fofr}) is asymptotically $\textmd{AdS}_4$.
\begin{figure}
\centering
\resizebox{0.6\textwidth}{!}{
\includegraphics{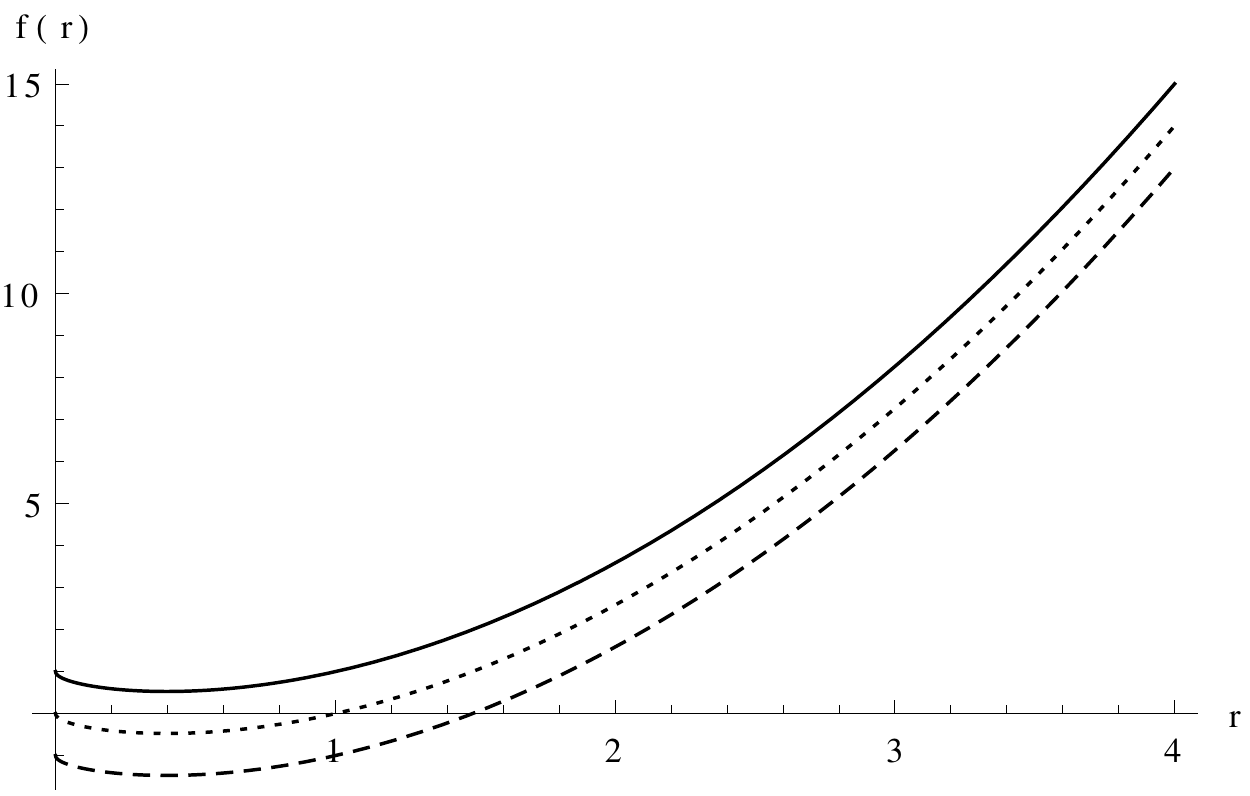}
}
\caption{Plots of $f(r)$ vs $r$ for $k=0$ (solid line), $k=1$ (dotted line) and $k=-1$ (dashed line) with $\Lambda_W=-1$, $a=1$, $M=1$}
\label{fofr_fig}     
\end{figure}
From the first law of black hole mechanics \cite{bardeen}, when a black hole undergoes a change from a stationary state to another, then the change in mass 
of the black hole is given by,
\begin{equation}
 dM=\frac{\kappa}{8 \pi} \, dA+\Omega\,dJ+\Phi \, dQ.
 \label{blackhole_first}
\end{equation}
Comparing this with the first law of thermodynamics,
\begin{equation}
 dM=T\,dS+\Omega\,dJ+\Phi \, dQ,
\end{equation}
one can easily establish the analogy between black hole mechanics and the first law of thermodynamics .
We know that Ho\v{r}ava-Lifshitz theory does not have possess the full diffeomorphism invariance of general relativity but only a 
subset in the form of local Galilean invariance. This subset is manifest in the Arnowitt, Deser and Misner (ADM) slicing. Here we have considered the
ADM decomposition of the four dimensional metric. Then for a non-rotating uncharged black hole, 
the entropy can be written as \cite{Kofinas, Padilla},
\begin{equation}
 S=\int \frac{dM}{T}= \int \frac{1}{T_H} \frac{\partial H}{\partial r_h} dr_h,
 \label{entropy_relation}
\end{equation}
where $H$ denotes the enthalpy and $r_h$ denotes the horizon radius. And the Hawking temperature can be determined from,
\begin{figure}
\centering
\resizebox{0.6\textwidth}{!}{
\includegraphics{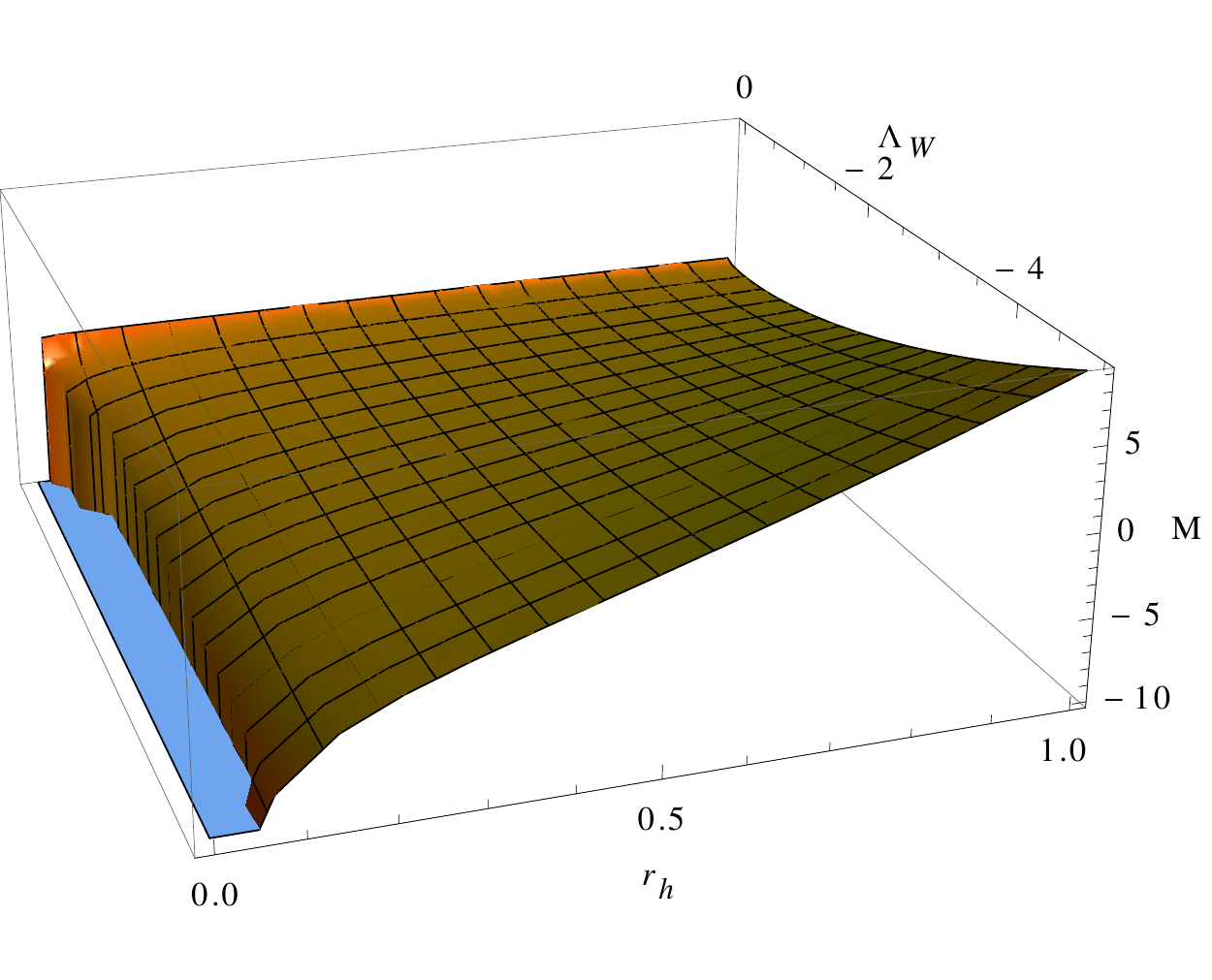}
}
\caption{3D Plots of Mass vs $r_h$}
\label{mass_fig} 
\end{figure}
\begin{equation}
 T_H=\frac{\kappa}{2 \pi}=\frac{1}{4\pi} f'(r)\vline_{r=r_h} ,
 \label{kappa_relation}
\end{equation}
and,
\begin{equation}
 T_H= \left( \frac{\partial H}{\partial S} \right)_{P},
 \label{temperature_relation}
\end{equation}
\begin{figure}
\centering
\resizebox{0.6\columnwidth}{!}{
\includegraphics{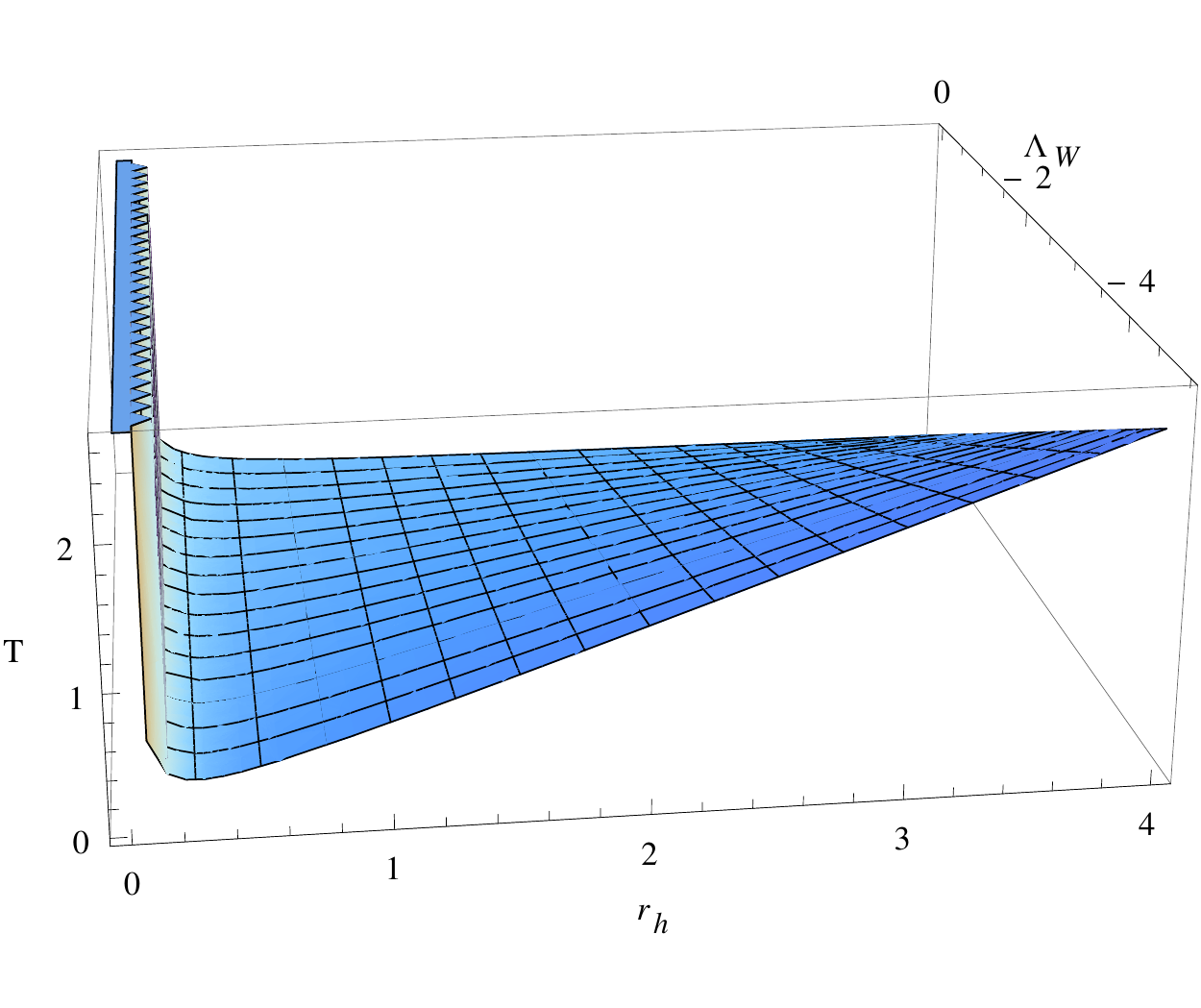}
}
\caption{3D Plots of temperature}
\label{temp_fig}
\end{figure}
where $P$ is the pressure and it is related to the Hawking temperature and entropy as,
\begin{equation}
 P=\frac{1}{2} T_{H} S.
 \label{pressure_relation}
\end{equation}
Hence the volume of the black hole is given by,
\begin{equation}
 V=\left( \frac{\partial H}{\partial P} \right)_S .
 \label{volume_relation}
\end{equation}
The heat capacity at constant pressure and at constant volume can be obtained respectively as,
\begin{equation}
C_P = T \left(\frac{\partial S}{ \partial T} \right) _P,
\label{cp_relation}
\end{equation}
and,
\begin{equation}
 C_V = C_P + V \frac{\partial P}{ \partial T}.
 \label{cv_relation}
\end{equation}
Using these relations we can calculate the thermodynamic quantities of the LMP black holes in arbitrary space curvature. In the cases of spherical$(k=1)$
and flat spaces $(k=0)$, detailed studies are done in \cite{sadeghi}. And it is noted that, in both cases the black hole doesn't show any kind of phase transition behaviors. 
In this paper we are interested in the LMP black hole solution in Hyperbolic space $(k=-1)$.
In this case, (\ref{fofr}) can be reduced to,
\begin{equation}
 f(r)= -1- \Lambda_W r^2 - \mathcal{A} \sqrt{\frac{r}{-\Lambda_W}}.
 \label{fofr_hyperbolic}
\end{equation}
The event horizon can be obtained from $f(r_h)=0$, and from that one can easily arrive at the black hole mass-event horizon radius relation as,
\begin{equation}
M=\frac{1}{a} \sqrt{\frac{-\Lambda_W}{r_h}} \left( -1 -\Lambda_W r^2_h \right) .
\label{mass_hyperbolic}
\end{equation}
In fig.(\ref{mass_fig}) we draw the 3D plot of variation of black hole mass with respect to black hole horizon radius for varying cosmological constant 
term($\Lambda_W$). From this figure it can be easily seen that the black hole mass increases with increase in the magnitude of  the cosmological
constant. It is interesting  to note from this figure that the black hole vanishes for lower values of $S$. Considering the numerical values, say
$a=1$ and $\Lambda_W=-1$ the black hole vanishes at $r_h=1$, but when $\Lambda_W=-2$ the black hole 
vanishes at another value of horizon radius, $r_h=0.5$, and so on. 

Black hole entropy can be obtained from (\ref{entropy_relation}) as,
\begin{equation}
 S=\frac{8 \pi \sqrt{-\Lambda_W }}{a} .
 \label{entropy_hyperbolic}
\end{equation}
From (\ref{kappa_relation}) we can derive the Hawking temperature as,
\begin{equation}
 T_H= \frac{\left( 1-3 \Lambda_W r^2_h \right)}{8 \pi r_h}.
 \label{temperature_hyperbolic}
\end{equation}
3D plot of Hawking temperature with respect to black hole horizon radius for varying cosmological constant term is depicted
in fig.(\ref{temp_fig}). Here also the temperature increases with the magnitude of  the cosmological constant. 
From (\ref{pressure_relation}), black hole pressure can be found as
\begin{equation}
 P= \frac{21 r_h \left( 1- 3 \Lambda_W r^2_h \right)}{64} .
 \label{pressure_hyperbolic}
\end{equation}
From the above expression, it is obvious that for any negative value of the cosmological constant term $\Lambda_W$, the pressure is found to 
be positive. Using (\ref{volume_relation}), black hole volume is given by,
\begin{equation}
 V= \frac{32 \sqrt{-\Lambda_W}}{21 a r^{\frac{3}{2}}_h} \frac{\left( 1- 3 \Lambda_W r^2_h \right)}{\left( 1- 9 \Lambda_W r^2_h \right)} .
 \label{volume_hyperbolic}
\end{equation}
 Using (\ref{cp_relation}) and (\ref{cv_relation}), the heat capacity at constant pressure and at constant volume are respectively determined as, 
 \begin{equation}
  C_P= \frac{4 \pi \sqrt{-\Lambda_W r_h}}{a} \frac{\left( 3 \Lambda_W r^2_h -1 \right)}{\left( 3 \Lambda_W r^2_h +1 \right)}
  \label{cp_hyperbolic}
 \end{equation}
and
 \begin{equation}
  C_V= \frac{8 \pi \sqrt{-\Lambda_W r_h}}{a} \frac{\left( 3 \Lambda_W r^2_h -1 \right)}{\left( 3 \Lambda_W r^2_h +1 \right)}
  \label{cv_hyperbolic}
 \end{equation}
 \begin{figure}
\centering
\resizebox{0.6\textwidth}{!}{
\includegraphics{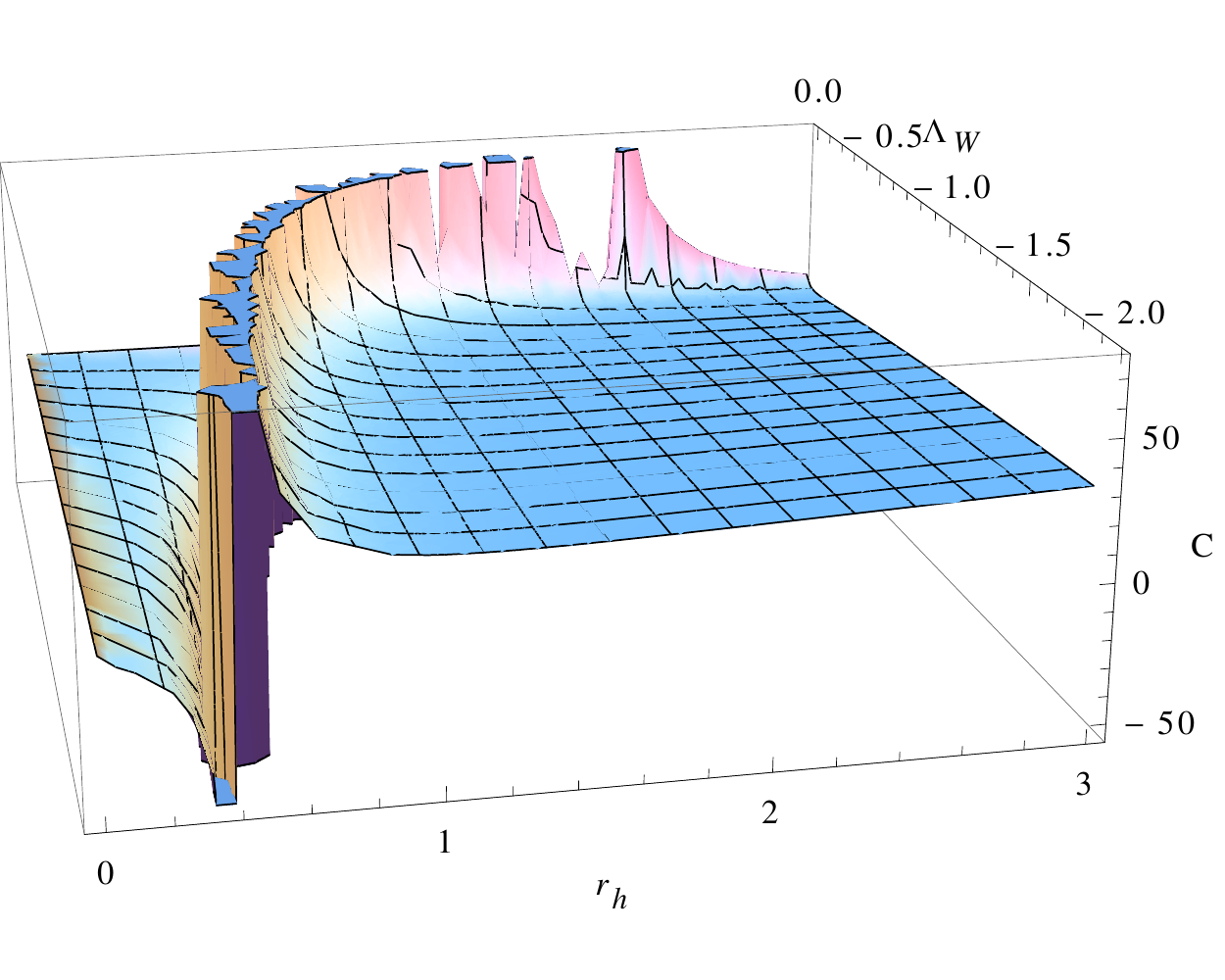}
}
\caption{3D Plots of heat capacity}
\label{spec_fig}  
\end{figure}
In fig.(\ref{spec_fig}) we draw the 3D variations of heat capacity with respect to horizon radius and for varying cosmological constant term.
From the 
figure, it is evident that the black hole has both positive and negative values in certain parametric regions. It is also clear from
the figure that, heat capacity has a divergent point. According to Davies \cite{davies}, second order phase transitions takes place at those points
where the heat capacity diverges. So LMP black hole undergoes a phase transition in this case. By investigating the free energy of the black hole
we can get a clear picture of the phase transition.
Free energy of the black hole is given by,
\begin{equation}
 F=M-TS.
 \label{freeenrgy_relation}
\end{equation}
Using (\ref{mass_hyperbolic}), (\ref{entropy_hyperbolic}) and (\ref{temperature_hyperbolic})  free energy of LMP black hole is
obtained as,
\begin{equation}
 F=\frac{1}{a} \sqrt{\frac{-\Lambda_W}{r_h}} \left( \frac{2(1+\Lambda_W r^2_h)(1-9\Lambda_W r^2_h)+(3 \Lambda_W r^2_h -1)^2}{2(9 \Lambda_W r^2_h -1)}            \right).
 \label{freeenrgy_hyperbolic}
\end{equation}
\begin{figure}
\centering
\resizebox{0.6\textwidth}{!}{
\includegraphics{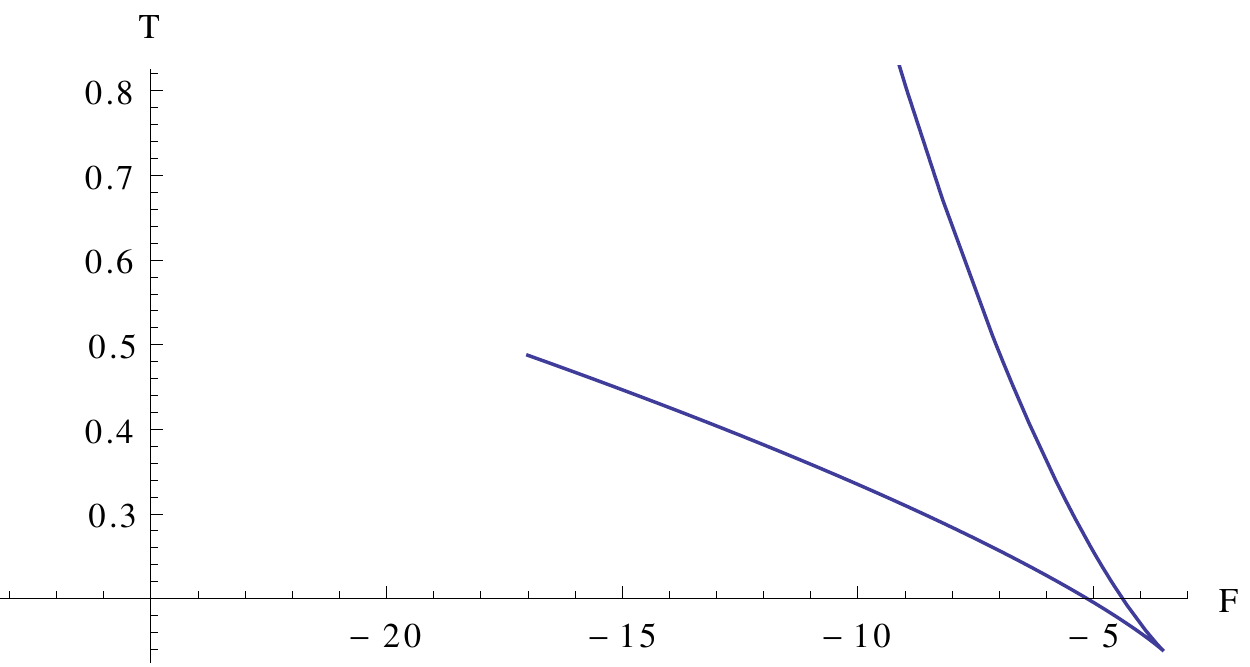}
}
\caption{parametric plot of free energy and temperature for $\Lambda_W=-1$,$a=1$}
\label{ft_fig}     
\end{figure}
Now one can plot the parametric variation of free energy and temperature using (\ref{freeenrgy_hyperbolic}) and (\ref{temperature_hyperbolic}). 
From fig (\ref{ft_fig}), it is evident that there is a cusp like double point. This indicates a second order phase transition. For one of the branches
the
free energy decreases and reaches the temperature which corresponds to the minimum free energy. There after, free energy increases with a different slope. 

\begin{figure}
\centering
\resizebox{0.6\textwidth}{!}{
\includegraphics{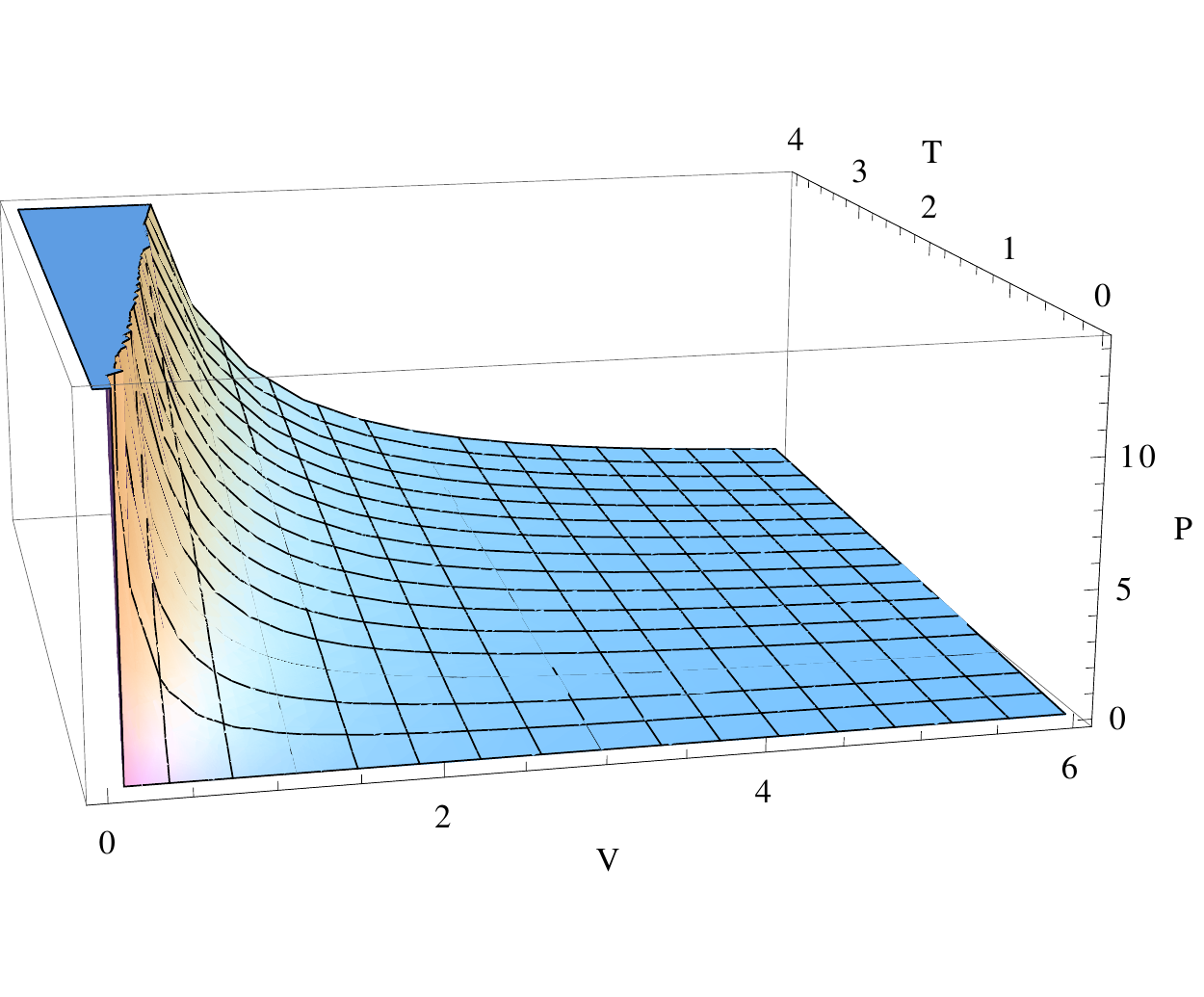}
}
\caption{Isotherm $P-V$ diagram}
\label{pv_fig}      
\end{figure}

An equation of state in general is a thermodynamic equation describing the state of matter under a given set of physical conditions.
It is a constitutive equation which provides  a mathematical relationship between two or more state functions associated with the matter,
such as its temperature, pressure, volume, or internal energy and 
black hole equation of state can be written from (\ref{pressure_hyperbolic}), (\ref{volume_hyperbolic}) and (\ref{temperature_hyperbolic}) as
\begin{equation}
 P V^{\frac{4}{3}} = \frac{4\pi T}{3} \sqrt{\frac{32}{63}} \frac{-\Lambda_W}{a^2}.
 \label{pvequation}
\end{equation}
Here, $P$ is the the pressure, $V$ is the thermodynamic volume and $T$ is the black hole temperature.
Now, we plot the isotherm $P-V$ diagram in fig (\ref{pv_fig}). From (\ref{pvequation}) and the corresponding figure (fig (\ref{pv_fig})) we can conclude that
the behaviour is "ideal gas like". Hence no critical point can be found and there would be no $P-V$ criticality.

\section{Geometrothermodynamics}
\label{gtd}
In order to introduce the idea of differential geometry to thermodynamics, according to \cite{quevedo1,quevedo2,quevedo3}, we have to define a $(2n+1)$ dimensional thermodynamic phase 
space $\mathcal{T}$. It can be coordinated by the set $Z^A=\{ \Phi, E^a ,I^a \}$, where $\Phi $ represents the thermodynamic potential and $E^a$ and $I^a$ 
represent extensive and intensive thermodynamic variables respectively. The phase space is provided with the Gibbs 1-form
$\Theta = d\Phi- \delta_{ab} I^a dE^b$, satisfying the condition, $\Theta \wedge (d \Theta)^{n} \neq 0$. Consider a Riemannian metric
$G$ on $\mathcal{T}$, which must be invariant with respect to Legendre transformations. Then the Riemannian contact manifold can be defined
as the set $(\mathcal{T},\Theta,G)$. Then the equilibrium manifold can be written as a sub manifold of $\mathcal{T}$, i.e., $\mathcal{E} \subset \mathcal{T}$.
This sub manifold satisfies the condition $\varphi^*(\Theta) = 0$ (where $\varphi : \mathcal{E} \rightarrow \mathcal{T}$), known as the pull back condition \cite{jackle}. 
Then the non-degenerate metric $G$ and the thermodynamic metric $g$ can be written as,
\begin{equation}
 G=(d\Phi - \delta_{ab} I^a d E^b)^2 +(\delta_{ab} E^a I^b)(\eta_{cd} d E^c d I^d),
\end{equation}
and,
\begin{equation}
  g^Q=\varphi^*(G)=\left(E^{c}\frac{\partial{\Phi}}{\partial{E^{c}}}\right)
\left(\eta_{ab}\delta^{bc}\frac{\partial^{2}\Phi}{\partial {E^{c}}\partial{E^{d}}} dE^a dE^d \right),
\label{quevedo metric}
\end{equation}
with $\eta_{ab}$=diag(-1,1,1,..,1) and this metric is Legendre invariant because of the invariance of the Gibbs 1-form.

Now we will introduce this idea of Geometrothermodynamics in to the LMP black hole system to study whether the black hole exhibits a
phase transition or not.
For this, we will consider a 5-dimensional thermodynamic phase space $\mathcal{T}$ constituted by extensive variables $S$ and $a$ and
the corresponding intensive 
variables $T$ and $A$. Thus the fundamental 1-form defined on $\mathcal{T}$ can be written as,
\begin{equation}
 d\Theta=dM-T\, dS-a\, dA
\end{equation}
Now the Quevedo metric \cite{quevedo1} is given by,
\[
         g=(SM_{S}+aM_{a})
            \left[ {\begin{array}{cc}
             -M_{SS} & 0  \\
             0 & M_{aa}  \\
             \end{array} } \right].
        \]
Then the Legendre invariant scalar curvature corresponding to the above metric is given by,
\begin{equation}
R=\frac{48 a^2 r^3 \left(35 r^6 \Lambda _W^3-5 r^4 \Lambda _W^2+9 r^2 \Lambda _W+9\right)}{\left(r^2 \Lambda _W+3\right){}^2 \left(3
   r^2 \Lambda _W+1\right){}^2 \left(5 r^2 \Lambda _W-3\right){}^3}.
 \label{curvature_scalar}
\end{equation}
\begin{figure}
\centering
\resizebox{0.6\textwidth}{!}{
\includegraphics{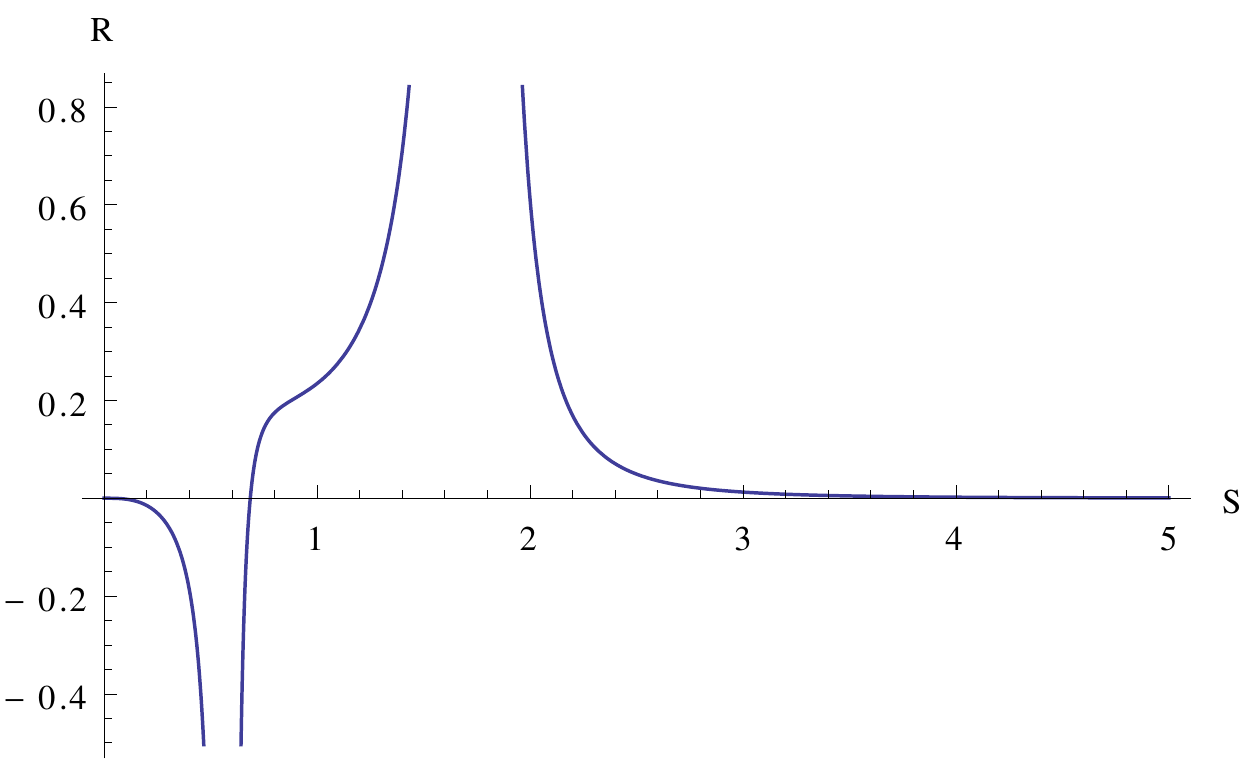}
}
\caption{variation of scalar curvature with horizon radius for $\Lambda_W=-1$,$a=1$}
\label{curvature_fig}      
\end{figure}
We have plotted the variation of scalar curvature with horizon radius in fig(\ref{curvature_fig}). From this figure as well as from the above equation, it can be confirmed that
this scalar curvature diverges at the same point where the heat capacity diverges. Hence the Geometrothermodynamics exactly reproduces the phase transition
structure of the LMP black hole.

\section{Analytical check of classical Ehrenfest equations}
\label{ehren}

Infinite discontinuity in the heat capacity of the black hole does not always indicate a second order phase transition, but it suggests the possibility of a 
higher order phase transition. In classical thermodynamics, one can confirm the first order phase transition by utilizing Clausius-Clapeyron equations. 
Similarly a second order transition can be confirmed by checking whether it satisfies Ehrenfest equations or not. The original expressions of Ehrenfest equations in classical
thermodynamics are given by,
\begin{equation}
 \left( \frac{\partial P}{\partial T} \right)_S = \frac{C_{P_{2}}-C_{P_{1}}}{VT (\alpha_2 - \alpha_1)} = \frac{\Delta C_P}{VT \Delta \alpha},
 \label{classical_ehrenfest_first}
\end{equation}
\begin{equation}
 \left( \frac{\partial P}{\partial T} \right)_V = \frac{\alpha_2 - \alpha_1}{\kappa_{T_{2}}-\kappa_{T_{1}}} = \frac{\Delta \alpha}{\Delta \kappa},
 \label{classical_ehrenfest_second}
\end{equation}
where $\alpha= \frac{1}{V} \left( \frac{\partial V}{\partial T} \right)_P $ is the volume expansion coefficient and 
$\kappa_T = -\frac{1}{V} \left( \frac{\partial V}{\partial P} \right)_T$ is the isothermal compressibility coefficient. Considering the analogy between
the thermodynamic variables and black hole parameters, where pressure $(P)$ is replaced by the negative of the electrostatic potential difference $(-\Phi)$,
and volume $(V)$ is replaced by charge of the black hole $(Q)$.  Thus for black hole thermodynamics, the two Ehrenfest equations
(\ref{classical_ehrenfest_first}) and (\ref{classical_ehrenfest_second}) become,
\begin{equation}
 -\left( \frac{\partial \Phi}{\partial T} \right)_S = \frac{1}{QT} \frac{C_{\Phi_{2}}-C_{\Phi_{1}}}{(\alpha_2 - \alpha_1)} = \frac{\Delta C_\Phi}{QT \Delta \alpha},
 \label{blackhole_ehrenfest_first}
\end{equation}
\begin{equation}
 -\left( \frac{\partial \Phi}{\partial T} \right)_Q = \frac{\alpha_2 - \alpha_1}{\kappa_{T_{2}}-\kappa_{T_{1}}} = \frac{\Delta \alpha}{\Delta \kappa},
 \label{blackhole_ehrenfest_second}
\end{equation}
where $\alpha= \frac{1}{Q} \left( \frac{\partial Q}{\partial T} \right)_\Phi $ is the volume expansion coefficient and 
$\kappa_T = -\frac{1}{Q} \left( \frac{\partial Q}{\partial \Phi} \right)_T$ is the isothermal compressibility coefficient of the black hole system.
Here, in the above sets of equations, the subscripts 1 and 2 denote two distinct phases of the system. 

\begin{figure}
\centering
\resizebox{0.6\textwidth}{!}{
\includegraphics{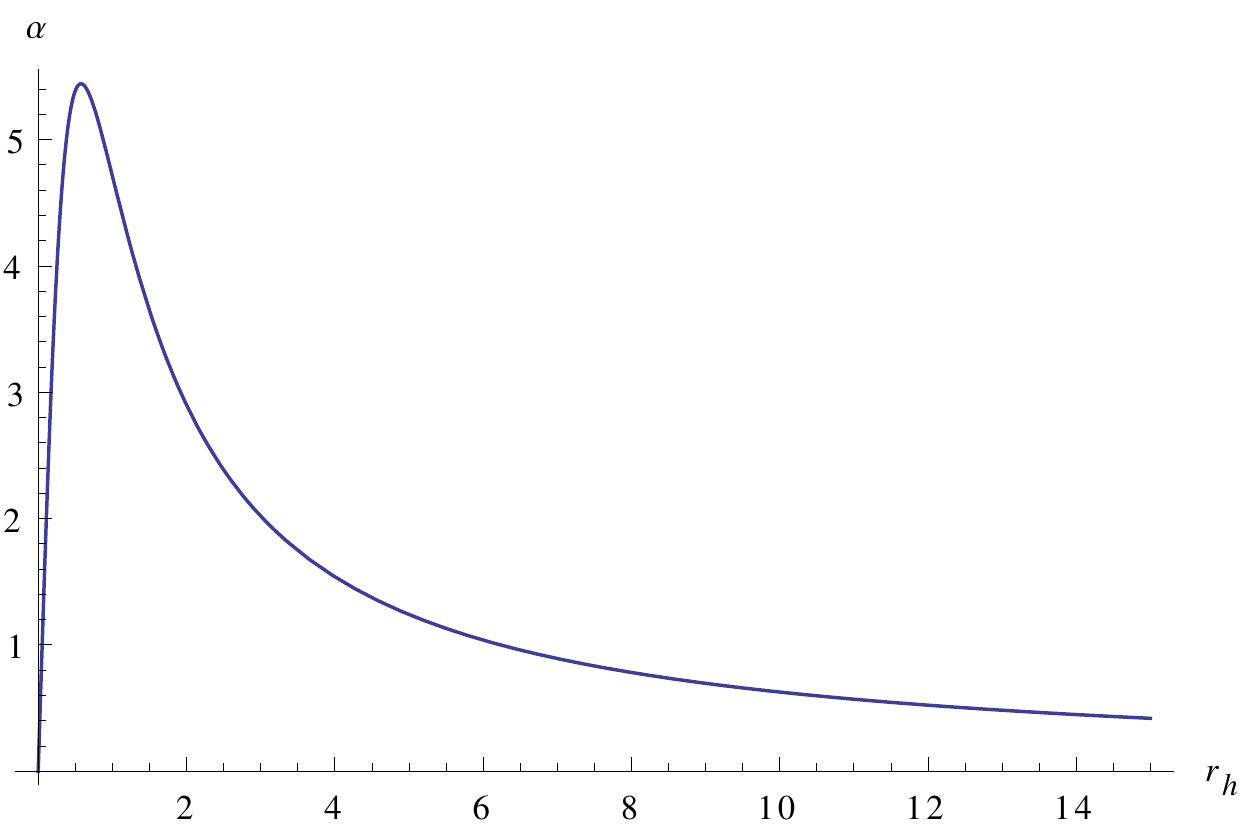}
}
\caption{variation of volume expansion coefficient with horizon radius for $\Lambda_W=-1$,$a=1$}
\label{alpha_fig}      
\end{figure}

\begin{figure}
\centering
\resizebox{0.6\textwidth}{!}{
\includegraphics{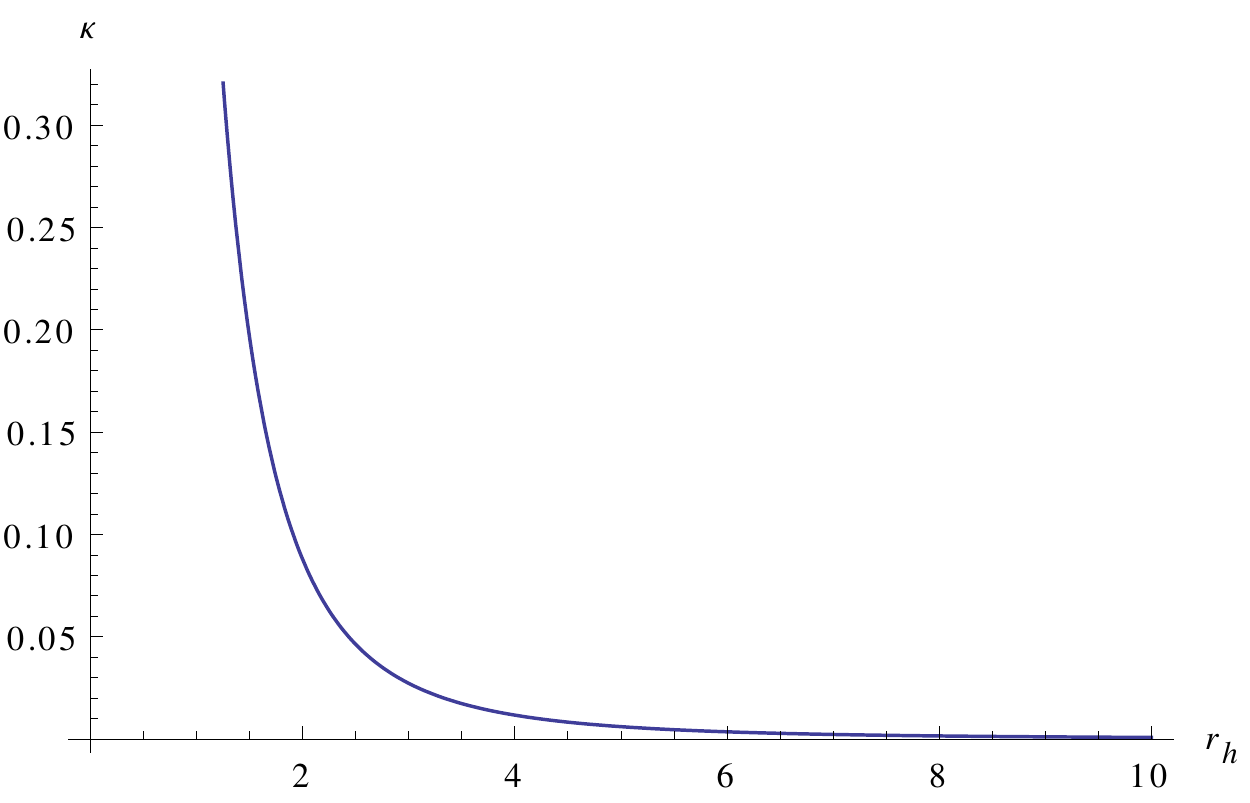}
}
\caption{variation of isothermal compressibility with horizon radius for $\Lambda_W=-1$,$a=1$}
\label{kappa_fig}      
\end{figure}

In this paper, rather than considering the black hole analogy of Ehrenfest equation, 
we will introduce the classical Ehrenfest equation directly in to the black hole system under consideration. 
Using (\ref{entropy_hyperbolic}), (\ref{temperature_hyperbolic}), (\ref{pressure_hyperbolic}) and (\ref{volume_hyperbolic}), we can arrive at the expressions 
of specific heat at constant pressure, volume expansion coefficient and isothermal compressibility coefficient respectively as,
\begin{equation}
  C_P= \frac{4 \pi \sqrt{-\Lambda_W r_h}}{a} \frac{\left( 3 \Lambda_W r^2_h -1 \right)}{\left( 3 \Lambda_W r^2_h +1 \right)},
  \label{cp_hyperbolic1}
\end{equation}
\begin{equation}
 \alpha=\frac{6 \pi r_h}{1-3r^2_h \Lambda_W},
 \label{volume expansion coefficient}
\end{equation}
and,
\begin{equation}
\kappa=\frac{16}{7 r_h} \frac{1}{1-3r^2_h \Lambda_W}.
\label{isothermal compressibility}
\end{equation}
From these relations, it is interesting to note that both volume expansion coefficient and isothermal compressibility coefficient have same factor in the 
denominator, which implies that both these parameters diverge at the same point. We have plotted the variation of these coefficients with respect to 
the horizon radius $(r_h)$ in figures (\ref{alpha_fig}) and(\ref{kappa_fig}) respectively.

Now we will investigate the nature of phase transition at the critical point of LMP black hole by doing the analytic check of 
classical Ehrenfest equations (\ref{classical_ehrenfest_first}) and (\ref{classical_ehrenfest_second}). The values of temperature, 
pressure and volume at the critical point are respectively given by,

\begin{equation}
T_c= \frac{\sqrt{3} \sqrt{\Lambda_W}}{4 \pi},
 \label{temperaturecritical}
\end{equation}
\begin{equation}
P_c=\frac{7 \sqrt{3} }{32 \sqrt{\Lambda_W}},
\label{pressurecritical}
\end{equation}
and,
\begin{equation}
 V_c=\frac{16 (-\Lambda_W )^{5/4}}{7 \sqrt[4]{3} a}.
 \label{volumecritical}
\end{equation}
Now let's check the validity of Ehrenfest equations at the critical point. 
From the definition of volume expansion coefficient $\alpha$, given by (\ref{volume expansion coefficient}), we obtain,
\begin{equation}
V \alpha = \left( \frac{\partial V}{\partial T} \right) _P= \left( \frac{\partial V}{\partial S} \right) _P \left( \frac{\partial S}{\partial T} \right) _P
=\left( \frac{\partial V}{\partial S} \right) _P \left( \frac{C_P}{T} \right)
\end{equation}
then, the R.H.S of first classical Ehrenfest equation (\ref{classical_ehrenfest_first}) becomes,
\begin{equation}
 \frac{\Delta C_P}{T V \Delta \alpha} = \left[ \left(  \frac{\partial S}{\partial V}   \right)_P  \right]_{r_{cri}},
\end{equation}
where $r_{cri}$ denotes the critical point. Applying the above equation to LMP black hole system, we obtain,
\begin{equation}
 \frac{\Delta C_P}{T V \Delta \alpha} = \frac{21 \pi}{24} \frac{1}{-\Lambda_W}.
 \label{rhs_1}
\end{equation}
Now the L.H.S of first classical Ehrenfest equation (\ref{classical_ehrenfest_first}) becomes,
\begin{equation}
 \left [ - \left( \frac{\partial P}{\partial T} \right)_S \right ]_{r_{cri}} = \frac{21 \pi}{24} \frac{1}{-\Lambda_W}.
 \label{lhs_1}
\end{equation}
From (\ref{lhs_1}) and (\ref{rhs_1}), we can arrive at the conclusion that both L.H.S and R.H.S of first Ehrenfest equation are in good agreement at the critical 
point $r_{cri}$. 
From (\ref{volume expansion coefficient}) and (\ref{isothermal compressibility}), using the thermodynamic identity,
\begin{equation}
 \left( \frac{\partial V}{\partial P} \right)_T \left( \frac{\partial P}{\partial T} \right)_V \left( \frac{\partial T}{\partial V} \right)_P = -1 ,
\end{equation}
we can obtain,
\begin{equation}
 V \kappa_T = - \left( \frac{\partial V}{\partial P} \right)_T=  \left( \frac{\partial T}{\partial P} \right)_V 
 \left( \frac{\partial V}{\partial T} \right)_P= \left( \frac{\partial T}{\partial P} \right)_V V \alpha .
\end{equation}
Now the R.H.S of (\ref{classical_ehrenfest_second}) can be obtained as,
\begin{equation}
 \frac{\Delta \alpha}{\Delta \kappa_T}=\left [  \left( \frac{\partial P}{\partial T} \right)_V \right ]_{r_{cri}}=\frac{21 \pi}{24} \frac{1}{(-\Lambda_W) ^{\frac{4}{3}}}.
\label{rhs_2}
 \end{equation}
Also the L.H.S of (\ref{classical_ehrenfest_second}) can be obtained as,
\begin{equation}
 \left [  \left( \frac{\partial P}{\partial T} \right)_V \right ]_{r_{cri}}=\frac{21 \pi}{24} \frac{1}{(-\Lambda_W) ^{\frac{4}{3}}}.
 \label{lhs_2}
\end{equation}
From (\ref{lhs_2}) and (\ref{rhs_2}), we can obtain the conclusion that second Ehrenfest equation is satisfied at the critical points. 
Hence both the Ehrenfest equations are in good agreement at the critical point. Using
(\ref{rhs_1}) and (\ref{rhs_2}), the Prigogine-Defay (PD) ratio is
\begin{equation}
 \Pi=\frac{\Delta C_P \Delta \kappa_T}{T V (\Delta \alpha)^2}=1 .
 \label{pd_ratio}
\end{equation}
This confirms that the phase transition of LMP black hole in Ho\v{r}ava-Lifshitz gravity is second order in nature. 

\section{Conclusions}
\label{Conclusions}
The complete thermodynamics and phase transition picture of LMP black holes in Ho\v{r}ava-Lifshitz gravity has been investigated using both thermodynamic 
geometry and Ehrenfest's scheme. We have systematically analyzed the thermodynamics and phase transition. From this thermodynamic study,
absence of any discontinuity in entropy-temperature relationship eliminates the presence of any first order transition. Then the heat capacity 
is found to be diverging, thereby indicating the presence of a phase transition. But the order of phase transition was not revealed. To further clarify the 
existence of phase transition, geometrothermodynamics is applied. In which the critical point where the heat capacity diverges coincides with the diverging 
point of Legendre invariant geometrothermodynamic scalar curvature. Hence GTD metric exactly reproduces the phase transition structure of LMP black hole and 
their corresponding thermodynamic interactions. Here we can conclude that the curvature scalar behaves in a similar way as that of the black hole system.
Then we have conducted a detailed  analytic check of classical Ehrenfest equations on LMP black hole system. 
From this we have found that the LMP black hole satisfies this equations, and hence the phase transition is second order in nature. The PD ratio found in these 
calculations further witnesses the second order phase transition, and confirms that there is no deviation from this order. 
Hence it will be possible to answer whether one can have a quantum field theory at a finite
temperature by studying the thermodynamic stability of the black hole, as evident from the specific heat. However, the
black hole configuration must be favourable over pure thermal radiation in anti-de Sitter space; that is, have dominant negative free energy. The present 
black hole solution satisfies these conditions. Hence from this study one can look forward for the implication on the dual field theory
which exists on the boundary of the anti-de Sitter space.

\section*{Acknowledgement}
The authors wish to thank referee for the useful comments and  UGC, New Delhi for financial support through a major research project
sanctioned to VCK. VCK also wishes to acknowledge Associateship of IUCAA, Pune, India.

\end{document}